# Discrete Holomorphic Parafermions in the Eight Vertex Model

M. Tanhayi-Ahari and S. Rouhani

Physics Department,
Sharif University of Technology
Tehran, PO BOX 11155-9161, Iran.
November 6, 2012

Abstract

We show that holomorphic Parafermions exist in the eight vertex model. This is done by extending the definition from the six vertex model to the eight vertex model utilizing a parameter redefinition. These Parafermions exist on the critical plane and integrable cases of the eight vertex model. We show that for the case of staggered eight vertex model, these Parafermions correspond to those of the Ashkin-Teller model. Furthermore, the loop representation of the eight vertex model enabled us to show a connection with the O($n$) model which is in agreement with the six vertex limit found as a special case of the O($n$) model.

1. Introduction

Smirnov has provided an elegant proof that critical percolation in the scaling limit is a Schramm-Loewner Evolution (SLE) and is related to a conformal field theory with vanishing central charge [1]. When applied to the Ising model his proof takes on an interesting shape [2] namely discrete holomorphicity. Indeed the proof that any statistical mechanical model is well defined in the limit of vanishing lattice spacing is hard enough, let alone requiring in addition that Cauchy-Riemann conditions should also hold. Such proofs for other SLE models have been too hard to come by [2]. However it is generally believed that the continuum limit of certain statistical mechanical models should correspond to SLEs. Or more precisely the interface separating different phases at criticality is a random curve with scaling properties of an SLE. One may wonder, is it necessary to go to the scaling limit to prove conformal invariance? In other words is it not possible to define conformal invariance for non-zero lattice spacing, and then be able to make correspondence with conformal invariance at non zero lattice spacing. This raises the deeper question; can one define discrete holomorphicity? A number of different approaches to discrete Cauchy-Riemann conditions have been offered. At criticality these definitions connect up with the concept of Parafermions [3] [4], which can be defined in terms of order and disorder variables. Parafermions are also a concrete theoretical tool see for example [5]. Interestingly there exists a connection between Parafermions and integrability [6]. It was observed in [6] that such Parafermions are indeed discretely holomorphic, but only at the integrable critical points [7] [8]. However note that the Parafermions of Ref [5] are continuous whereas here we are mainly discussing the discrete case. Other examples are the q-state Potts model [9], the $Z_N$ model [10], the O($n$) model [6], and the Ashkin-Teller model [11] which is directly related to the six vertex model. More specifically the six vertex model (XXZ spin chain) has been shown to be related to the generalized minimal unitary series of conformal field theory and the $Z_N$ Parafermion theories [12]. Since the eight vertex model is closely related to the six vertex model and is integrable (for a defined parameter set) the natural question arises as to why Parafermions have not been defined for the eight vertex model? In this paper we show that indeed it is possible to define Parafermions for the eight vertex model and there is an intimate relation with criticality and integrability. Furthermore, due to the connection between staggered eight vertex model and the Ashkin-Teller (AT) model we can show that our Parafermions correspond to those of the AT model in a special case. Furthermore the loop representation of the eight vertex model, is closely connected with the O($n$) loop model. In other words, eight vertex model could be considered as a special case of the O($n$) model. Moreover, using simple symmetry properties of the eight vertex model we are able to show a mapping of the O($n$) model to the eight vertex model.

This paper is organized as follows; in section 2, we briefly review some models which admit Parafermions, in section 3, we give evidence of Parafermions existing in the eight vertex model and also show some correspondences with previous results. Section 4 is devoted to a brief conclusion and section 5 is for acknowledgments.

2. Brief Review of Previous works

In this section we briefly review some models for which Parafermions have been found and are related to the eight vertex model. This review is by no means inclusive, in particular many definitions and approaches to discrete holomorphicity exist which we are not covering.

2.1. Discrete Holomorphicity

A graph G embedded in $R^2$, is composed of vertices at $z_i$ (complex coordinates) and edges $(ij)$ connecting the vertices. Obvious examples are the square lattice and triangulation of the plane. Let $F(z_{ij})$ be a complex valued function defined at midpoints $z_{ij}$ of the edges $(ij)$. The discrete version of contour integral can be stated as:

$$\sum_{(ij)\epsilon C} F(z_{ij})(z_j - z_i) = 0 \qquad . \qquad\qquad 1$$

where C is any closed contour in G, notice that Eq. 1 is a direct discretization of the contour integral over the complex plane and then Morera's theorem ensures that $F$ would be holomorphic function satisfying the Cauchy-Riemann equations if the contour integral vanishes [13]. For a square lattice Eq.1 reduces to:

$$F(z_{12}) + iF(z_{23}) + i^2 F(z_{34}) + i^3 F(z_{41}) = 0 \qquad , \qquad\qquad 2$$

for every plaquette of the square lattice. Equation 2 can be rewritten for the honeycomb lattice with a replacement of the roots of unity. Clearly Eq.1 massively reduces the degrees of freedom, but fails to completely determine F. One would hope that this definition of discrete holomorphicity could have as much power as its continuous counterpart however this is not true although a lot of work has been done in this direction [14].

## 2.2. Self-Dual Ising Model

The standard definition of the Parafermion for the Ising model is a product of nearby order and disorder variables. To do this, we need to have a dual description in the sense of Kramers-Wannier [15]. On the other hand in the models in which we don't have this duality the Parafermions defined in terms of the random curves of the model.

Now for the self-dual Ising model on the square lattice with the Hamiltonian

$$H = -\sum_{(ij)} J_{ij} S_i S_j \qquad . \qquad 3$$

The partition function is:

$$Z = \prod_{(ij)} \cosh(J_{ij})(1 + x_{ij} S_i S_j) \qquad . \qquad 4$$

Where $x_{ij} = \tanh(J_{ij})$. The disorder variable is defined on the dual vertex $\tilde{\imath}$, as follows

$$\mu(\tilde{\imath}) = \prod_{(ij) \text{from boundary}}^{\tilde{\imath}} \frac{1 - x_{ij} S_i S_j}{1 + x_{ij} S_i S_j} \qquad . \qquad 5$$

Inserting $\mu$ is equivalent to replacing $J_{ij} \rightarrow -J_{ij}$ in every edge $(ij)$ crossed by the string connecting $\tilde{\imath}$ to the boundary. Now we can express the Parafermion on the mid-point of the connecting line from j to $\tilde{\imath}$ which is the midpoint of the covering lattice. Which covering lattice is the union of the lattice and it's dual:

$$\psi_s(j\tilde{\imath}) = S_j \mu_{\tilde{\imath}} e^{-is\theta} \qquad . \qquad 6$$

Where $\theta$ is the angle that $(j\tilde{\imath})$ makes with the positive x-axis (a fixed axis in plane) which varies in the interval $(-\pi, \pi)$, and $s$ is known as the spin of the Parafermion, in this case we have $s = 1/2$.

For homogenous Ising model where $x_{ij} \rightarrow x$, and for two neighboring dual points $\tilde{1}$ and $\tilde{2}$ in a typical elementary square of the lattice, it is easy to see that:

$$\mu_{\tilde{1}} = \frac{1 - x S_1 S_2}{1 + x S_1 S_2} \mu_{\tilde{2}} \qquad . \qquad 7$$

For special value of $x = \sqrt{2} - 1$ and $s = 1/2$, we obtain the discrete form of the contour integral, resulting in the form of Parafermion as a product of the order and disorder variable with a phase factor.

For the continuum critical Ising model, the underlying theory is the free massless real fermion, the Majorana fermion [16], given by the action

$$S = \frac{1}{2\pi} \int d^2z \left( \psi \bar{\partial} \psi + \bar{\psi} \partial \bar{\psi} \right) \qquad . \qquad 8$$

The action has conformal symmetry which stems from the criticality of the massless model. For the above action from the direct calculation, the OPE of the energy-momentum tensor with itself is [17];

$$T(z)T(w) \sim \frac{1/4}{(z-w)^4} + \frac{2T(w)}{(z-w)^2} + \frac{\partial T(w)}{(z-w)} \qquad . \qquad 9$$

The important point of the above OPE is the determination of the central charge as $c = 1/2$ for the Ising model. The central charge of the model is related to the conformal spin as $c = 2s(5 - 8s)/(2s + 1)$ (see [18] and [19]), which for $c = 1/2$ gives $s = 1/2$, this is exactly the same spin value obtained by imposing discrete holomorphicity on the Parafermions of the Ising model.

2.3. Self-dual Potts model

The Q-state Potts model on the square lattice is defined by spin variables $S_i = \{1, \dots, Q\}$, with the partition function:

$$Z = \sum_{\{S\}} \exp\left( \sum_{<i,j>} J \delta(S_i, S_j) \right) = \sum_{\{S\}} \prod_{<i,j>} (1 + u\delta(S_i, S_j)) \qquad . \qquad 10$$

Where $u = e^J - 1$ and $\delta(S_i, S_j)$ is equal to unity if $S_i = S_j$ and otherwise zero. Expanding the product one gets Fortuin-Kasteleyn (FK) representation [20], [9].

$$Z = \sum_G u^b Q^c \qquad . \qquad 11$$

Here G is any sub graph of the original domain, consisting of all the sites and some bonds placed arbitrary on the lattice edges, and b is the number of bonds in G, and c is the number of clusters of the connected sites.

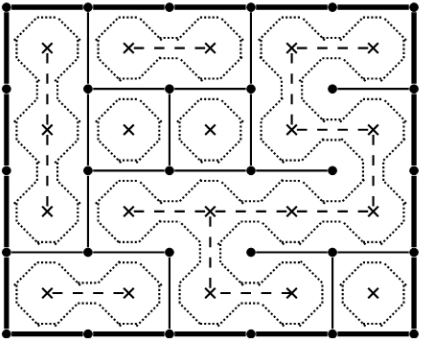

Figure 1. A configuration in the FK representation of the Potts model with wired boundary conditions (thick lines), dots and crosses represents sites of the original and dual lattice, respectively.

By imposing self-duality one gets, extracting the critical value, $u_c = \sqrt{Q}$ which states that weights of the FK clusters at criticality are equivalent to counting each loop on the medial lattice with a fugacity $\sqrt{Q}$. Now we can define a Parafermion on the edges of the covering lattice "$e$", which consists of both the original and the dual lattice,

$$\psi_s(e) = S(e)\,\mu(e)e^{-is\theta} \qquad . \qquad 12$$

Here $S(e)$ and $\mu(e)$ are the order (spin) and disorder variables on either side of the edge $e$. The correlation function,

$$< \psi_s(e_1)\psi_s(e_2)\ldots\psi_s(e_n) > \qquad , \qquad 13$$

satisfies the discrete version of the contour integral if we choose the values of $s$ and $Q$, as it is shown in [9], as below

$$\sqrt{Q} = 2\sin\frac{\pi s}{2} \qquad . \qquad 14$$

Consequently, $Q$ determines the spin of the Parafermion of the Potts model.

2.3. The O(n) model

For the case of O(n) model, the observable $F_s(z)$ is defined on the midpoints of the plaquette edges as follows;

$$F_s(z) = \sum_{G\in\Gamma(0,z)} P(G)e^{-is\theta} \qquad . \qquad 15$$

The sum is over graphs where there is an oriented open path going from 0 to z. Associating with six vertices of the O(n) loop model, the following Boltzmann weights:

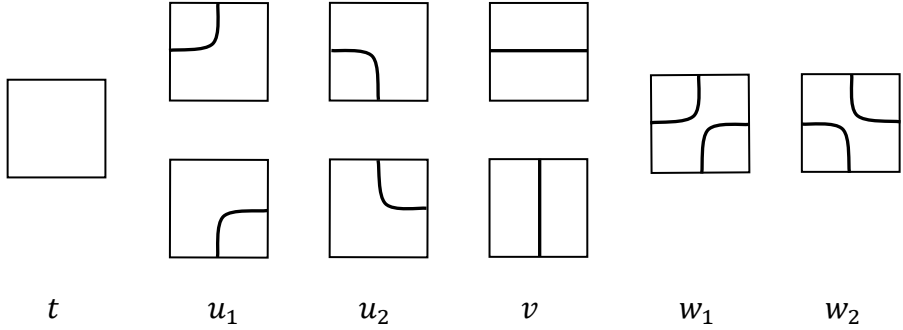

Figure 2. Vertices of the O(n) loop model on the square lattice, each closed loop has a weight $n$.

Summing over all internal configurations of paths satisfying Eq.1, we extract some linear equations for six unknown Boltzmann weights. Then the Parafermion weights can be obtained [6];

$$t = -\sin\left(2\varphi - \frac{3\eta}{2}\right) + \sin\frac{5\eta}{2} - \sin\frac{3\eta}{2} + \sin\frac{\eta}{2} \quad , \quad \text{16-I}$$

$$u_1 = -2\sin\eta \cos\left(\frac{3\eta}{2} - \varphi\right) \quad , \quad \text{16-II}$$

$$u_2 = -2\sin\eta \sin\varphi \quad , \quad \text{16-III}$$

$$v = -2\sin\varphi \cos\left(\frac{3\eta}{2} - \varphi\right) \quad , \quad \text{16-IV}$$

$$w_1 = -2\sin(\varphi - \eta)\cos\left(\frac{3\eta}{2} - \varphi\right) \quad , \quad \text{16-V}$$

$$w_2 = 2\sin(\varphi)\cos\left(\frac{\eta}{2} - \varphi\right) \quad . \quad \text{16-VI}$$

Where $s = \frac{3\eta}{2\pi} - \frac{1}{2}$ , $-\pi \leq \eta \leq \pi$ and $n = -2\cos 2\eta$.

We focus on the non-trivial solution of the vanishing case $v = 0$ with $n=1$, in which we get:

$$t = \sin\pi s, \; u_1 = \sin(\varphi - \pi s), \; u_2 = \sin\varphi, \; w_1 + w_2 = \sin\pi s \quad . \quad 17$$

This model can be mapped onto the six vertex model by choosing:

$$a = \omega_1 = \omega_2 = u_1, \; b = \omega_3 = \omega_4 = u_2, \; c = \omega_5 = \omega_6 = w_1 + w_2 \quad . \quad 18$$

The corresponding vertices of the six vertex model and plaquettes of the O(*n*) model are given in Fig 3.

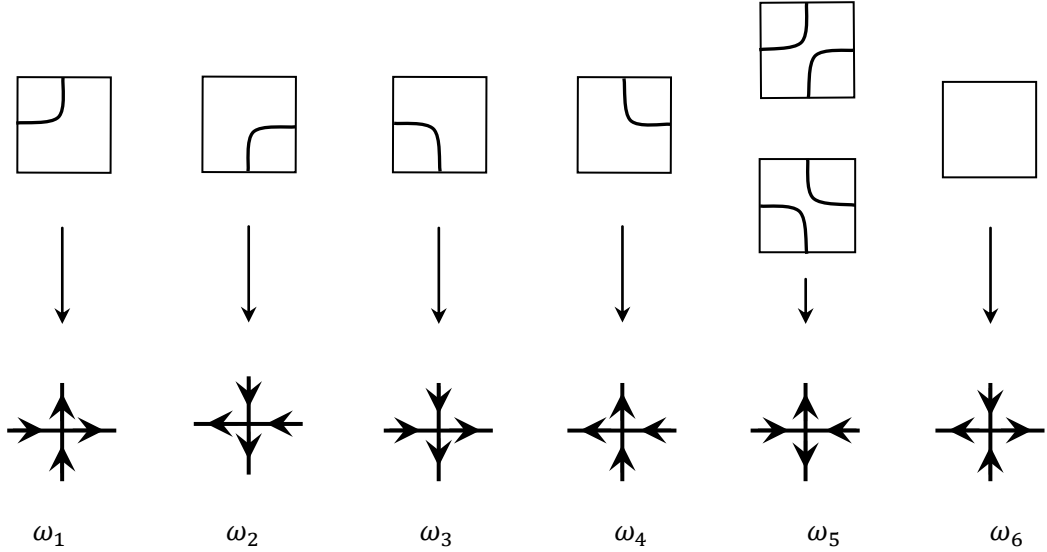

Figure 3. Mapping of the O(*n*) model ($v = 0, n = 1$) onto the six vertex model.

We use this connection to find the Parafermions of the eight vertex model.

2.4 The Ashkin-Teller model

In reference [11] discrete Parafermions all along the critical line of the AT model are found. These Parafermions are defined in the loop formulation of the AT model and the key idea is to use the mapping to O(*n*) model with ($v = 0, n = 1$). Therefore a six vertex model, with weights given by;

$$a, b, c = \sin(\varphi - \pi s), \sin \varphi, \sin \pi s \qquad , \qquad 19$$

exhibits a discrete holomorphic Parafermion $F_s(z)$, and $\varphi$ is related to spin and angel of the rhombi, $\alpha$, in O(n) loop model which is linearly depended on spectral parameter.

$$\varphi = (s + 1)\alpha \qquad . \qquad 20$$

Here $a, b, c$, are exactly the integrable weights of the six vertex model. This paves the way for the eight vertex model.

## 3 The eight vertex model

The eight vertex model has been introduced as a model for (anti)ferroelectrics, as a generalized ice-type model [21], [22]. On every edge of the square lattice place an arrow with the restriction that an even number of arrows point into every vertex. Assign to every vertex a Boltzmann weight $\omega_i$, Fig. 4.

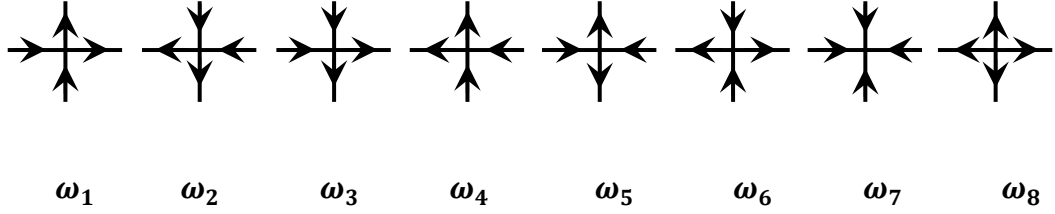

$\omega_1 \quad \omega_2 \quad \omega_3 \quad \omega_4 \quad \omega_5 \quad \omega_6 \quad \omega_7 \quad \omega_8$

Figure 4. Boltzmann weights of the eight vertex model.

Regarding the arrows as electric dipoles; the symmetric Anzats has the following weights:

$$a = \omega_1 = \omega_2 \quad b = \omega_3 = \omega_4 \quad c = \omega_5 = \omega_6 \quad d = \omega_7 = \omega_8 \quad . \quad\quad 21$$

These equalities mean that no external electric field is applied. The symmetric eight vertex model has been solved by Baxter [23], and the integrable weights are obtained by solving the star-triangle or Yang-Baxter equations. The parameterized integrable weights are as follows:

$$a:b:c:d = \operatorname{snh}(\lambda - u):\operatorname{snh}(u):\operatorname{snh}(\lambda):k\operatorname{snh}(\lambda - u)\operatorname{snh}(u)\operatorname{snh}(\lambda) \quad . \quad\quad 22$$

Where $\operatorname{snh}(u)$ is given in terms of the elliptic function with modulus k. Define $\Delta$ and $\Gamma$ as:

$$\Delta = \frac{a^2 + b^2 - c^2 - d^2}{2(ab + cd)} \quad , \quad \Gamma = \frac{ab - cd}{ab + cd} \quad . \quad\quad 23$$

We can change the Boltzmann weights ($a, b, c, d$ replaced with $a', b', c', d'$) and the model remains integrable as long as [24]:

$$\Delta = \Delta' \quad , \quad \Gamma = \Gamma' \quad . \quad\quad 24$$

The critical weights of the model can be obtained setting $k = 1$ in Eq. 22, or when the weights lie on surfaces like

$$a = b + c + d, \quad \text{or} \quad b = a + d + c, \quad \text{or} \quad c = a + b + d, \quad \text{or} \quad d = a + b + c \quad . \qquad 25$$

Without loss of generality we choose the case $c = a + b + d$. Therefore, we get $\Delta = -1$ for the chosen critical surface. For the case of

$$a > 0, \ b > 0, \ d > 0, \qquad 26$$

the critical exponent varies continuously on the critical surface $c = a + b + d$ as follows

$$f \sim |T - T_c|^{\frac{\pi}{\mu}} \quad \text{where} \quad \cos \mu = \frac{ab-cd}{ab+cd} = \Gamma \quad . \qquad 27$$

The solution in the rest of the $(a, b, c, d)$ space can be constructed from symmetry relations (see Ref [25], [26])

$$Z(a,b;c,d) = Z(\pm a, \pm b, \pm c, \pm d) = Z(b,a;c,d) = Z(c,d;a,b) =$$

$$Z\left(\tfrac{1}{2}(a-b+c+d), \tfrac{1}{2}(-a+b+c+d); \tfrac{1}{2}(a+b+c-d), \tfrac{1}{2}(a+b-c+d)\right). \qquad 28$$

If we substitute the criticality condition, $c = a + b + d$, in to the last term of the partition function, we get

$$Z_c(a,b;c,d) = Z(a+d, b+d; a+b, 0) \quad . \qquad 29$$

This shows that we could map the critical eight vertex model onto a six vertex model only by using symmetry together with the criticality condition.

3.1 Holomorphic Parafermions

As stated above, existence of discrete holomorphic Parafermions for the O(n) model has been shown, and the weights of Parafermions coincides with the integrable weights of the model given by Nienhuis in [27].

Considering equations 19 and 29, we can calculate the Parafermion weights for the eight vertex model as below:

$$a + d = \sin(\varphi - \pi s), \quad b + d = \sin\varphi, \quad a + b = \sin\pi s \qquad . \qquad 30$$

So

$$a = \frac{1}{2}(\sin(\varphi - \pi s) - \sin\varphi + \sin\pi s) \qquad , \qquad \text{31-I}$$

$$b = \frac{1}{2}(-\sin(\varphi - \pi s) + \sin\varphi + \sin(\pi s)) \qquad , \qquad \text{31-II}$$

$$c = \frac{1}{2}(\sin(\varphi - \pi s) + \sin\varphi + \sin(\pi s)) \qquad , \qquad \text{31-III}$$

$$d = \frac{1}{2}(\sin(\varphi - \pi s) + \sin\varphi - \sin(\pi s)) \qquad . \qquad \text{31-IV}$$

It is easy to see that with above weights we have,

$$\Delta = -1 \quad , \quad \Gamma = -\cos(\pi s) \qquad . \qquad 32$$

So it is consistent with the critical values of the $\Delta$ and $\Gamma$. It is useful to consider the special case of Ising model, $s = 1/2$, we have $\Gamma = 0$ hence we get $ab = cd$ or $\mu = \pi/2$, therefore Eq. 27 results in $f \sim |T - T_c|^2$ which is clear for critical exponent $\alpha = 0$ for 2d Ising model. Now we try to find a relation between our weights and the Ashkin-Teller weights. In AT model to each site $i$, two spin variables $s_i$ and $\sigma_i$ are associated. The interaction energy for the edge $(ij)$ is

$$E(ij) = -Js_i s_j - J' \sigma_i \sigma_j - J_4 s_i s_j \sigma_i \sigma_j - J_0 \qquad . \qquad 33$$

We use the dimensionless interaction coefficients [24],

$$K = \frac{J}{KT}, \quad K' = \frac{J'}{KT}, \quad K_4 = \frac{J_4}{KT}, \quad K_0 = \frac{J_0}{KT} \qquad . \qquad 34$$

Different weights of the model are given by;

$$w_0 = \exp(K + K' + K_4 + K_0), \quad w_1 = \exp(K - K' - K_4 + K_0),$$

$$w_2 = \exp(-K + K' - K_4 + K_0) , \quad w_3 = \exp(-K - K' + K_4 + K_0) . \qquad 35$$

The AT model is related to staggered eight vertex model on two sub-lattices A and B, with the following Boltzmann weights for sub-lattice A:

$$a = \frac{w_0 + w_1}{\sqrt{2}} , \quad b = \frac{w_2 - w_3}{\sqrt{2}} , \quad c = \frac{w_2 + w_3}{\sqrt{2}} , \quad d = \frac{w_0 - w_1}{\sqrt{2}} . \qquad 36$$

For sub-lattice B they are equal to $a, b, d, c$. Now we impose the criticality condition ($c = a + b + d$) on the above weights. Critical staggered eight vertex model gives:

$$w_0 = w_3 \text{ or } K + K' = 0 \qquad . \qquad 37$$

Now we calculate $w_i$, with the weights for the eight vertex model, after some simple algebra we get:

$$w_0 = \frac{\sin(\varphi - \pi s)}{\sqrt{2}} , \quad w_1 = \frac{\sin \pi s - \sin \varphi}{\sqrt{2}} , \quad w_2 = \frac{\sin \pi s + \sin \varphi}{\sqrt{2}} , \quad w_3 = \frac{\sin(\varphi - \pi s)}{\sqrt{2}} . \qquad 38$$

We see that the criticality condition still holds, that is $w_0 = w_3$, in the partition function of AT model we have terms like $\tanh 2K'$, $\frac{\exp(2K_4)}{\cosh 2K}$ using equations 39 and 36 we can see that:

$$\tanh 2K' = \frac{\exp(-K+K'-K_4+K_0) - \exp(K-K'-K_4+K_0)}{\exp(-K+K'-K_4+K_0) + \exp(K-K'-K_4+K_0)} = \frac{w_2 - w_1}{w_2 + w_1} = \frac{\sin \varphi}{\sin \pi s} , \qquad 39$$

$$\frac{\exp(2K_4)}{\cosh 2K} = \frac{\exp(K+K'+K_4+K_0)}{\frac{1}{2}(\exp(-K+K'-K_4+K_0) + \exp(K-K'-K_4+K_0))} = \frac{\sqrt{2} w_0}{\frac{1}{\sqrt{2}}(w_2 + w_1)} = \frac{\sin(\varphi - \pi s)}{\sin \pi s} . \qquad 40$$

These are the critical line of the AT model given in [11].

3.3 Connection with Ising spin model

One can define a simple map of the eight-vertex model onto the spin models [28], therefore we can simply relate our vertex model to SLE. In order to map the eight vertex model to Ising spin model, we can simply assign a spin placed at the interstitial points of the lattice obeying simple rules, an arrow to the right or upward corresponds to the situation in which the adjacent spins are parallel and leftward or downward arrow is a situation in which two adjacent spins are antiparallel, as in Fig.5. The eight vertex model is equivalent to two Ising models with nearest-neighbor coupling interaction $K^{\pm}$, via a four spin coupling term $\lambda$. The partition function of the model could be written as the following form:

$$Z = \sum_{\sigma = \pm 1} \prod_{j,k} A \, e^{K^+ \sigma_{j,k} \sigma_{j+1,k+1} + K^- \sigma_{j,k+1} \sigma_{j+1,k} + \lambda \sigma_{j,k} \sigma_{j+1,k+1} \sigma_{j,k+1} \sigma_{j+1,k}} . \qquad 41$$

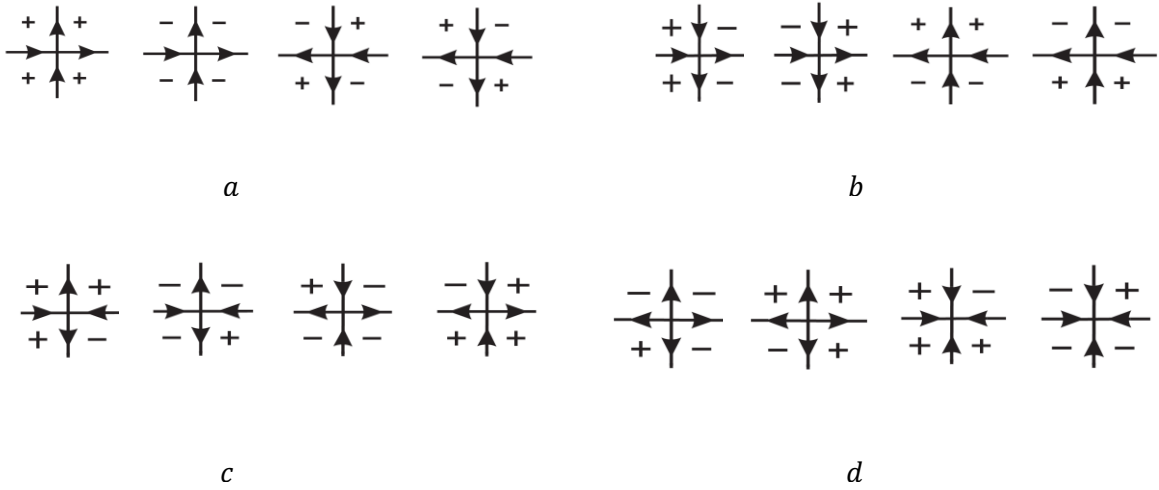

Figure5. The correspondence between the Ising spin and vertices of the eight vertex model.

The connection between four Boltzmann weight of the eight vertex model and this new spin model is:

$$a = A\, e^{K^+ + K^- + \lambda} \quad , \qquad\qquad 42\text{-I}$$

$$b = A\, e^{-K^+ - K^- + \lambda} \quad , \qquad\qquad 42\text{-II}$$

$$c = A\, e^{K^+ - K^- - \lambda} \quad , \qquad\qquad 42\text{-III}$$

$$d = A\, e^{-K^+ + K^- - \lambda} \quad . \qquad\qquad 42\text{-IV}$$

It is easy to see that the parameter $\Gamma$ is determined by four spin interaction coupling $\lambda$.

$$\Gamma = \tanh(2\lambda) \quad . \qquad\qquad 43$$

Notice that as we found above, the Parafermions spin which is related to $\Gamma$ as $\Gamma = -\cos(\pi s)$, is evidently specified by $\lambda$. The critical condition of the eight vertex model, $c = a + b + d$, could be written as below;

$$\frac{1+\Lambda}{1-\Lambda} = \frac{x-y}{1+xy} \quad , \qquad\qquad 44$$

or

$$\Lambda = \frac{x-y-xy-1}{x-y+xy+1} \quad . \qquad\qquad 45$$

where, $x = \tanh(K^+)$, $y = \tanh(K^-)$ and $\Lambda = \tanh(\lambda)$. Therefore at the critical point the four body coupling constant is determined by $K^+$ and $K^-$ only, thus given the two body interactions we would be able to determine the spin of the Parafermion.

3.4 Paths in the eight vertex model and connection with O($n$) model

Following Baxter's method to associate line configuration to six vertex model in which up and right arrows are taken to be thick lines and the down and left wards arrows to be light lines (alternatively one could think of two colors). Applying this method to the eight vertex model, gives a line configuration corresponding to the eight vertex model as below:

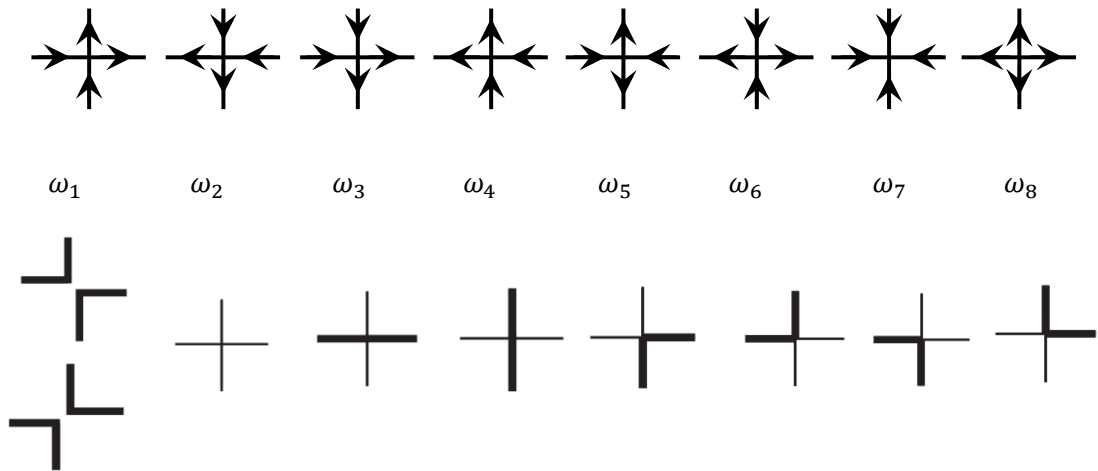

Figure 6. Line configuration associated with each vertex of the eight vertex model.

Now, in the eight vertex model any allowed configuration of the vertices could be mapped to a loop model constructed by the above line configuration. In the line configuration, any thick line must join to another thick line and this ensures we have a loop model, which the spin flips in crossing thick lines. The line configuration constructed above gives a hint of connection between eight vertex model and the O($n$) model. It is obvious that the line configuration in figure 6 has every line configuration of the O($n$) model shown in figure 2, in the eight vertex case this corresponds to zero field eight vertex model which is the case we are dealing with in this paper. The correspondence could be

$$t = w_1 + w_2 = a, \quad v = b, \quad u_1 = c, \quad u_2 = d,$$

$$Z(u_1, u_2, t = w_1 + w_2, v) = Z(c, d, a, b) \qquad . \qquad 46$$

Where $a, b, c, d$ are the Boltzmann weights for the eight vertex model. By symmetry properties of the eight vertex model, the above correspondence could be related to the six vertex model by permutation:

$$Z(c, d, a, b) = Z(a, b, c, d) \qquad . \qquad 47$$

If we take $v = 0$ we get the six vertex model. Moreover, we must take $n = 1$ in addition to $t = w_1 + w_2$ to have the connection. Therefore we can take any zero field eight vertex model as a special case of the O($n$= 1) model with $t = w_1 + w_2$. The definition of the Parafermion is in terms of the random curves of the model, like O($n$) model's Parafermion, would be sum over all paths starting from 0 to a special point z.

4. Conclusion

In this paper we have extended the notion of holomorphic Parafermions to the eight vertex model, by utilizing a twisted mapping between the six vertex and the eight vertex model. Furthermore we observe that both cases are connected with the O($n$) model. What is intriguing is that like all other cases, existence of the Parafermions is intimately connected with integrability. An insight into this is offered in ref [7] . As long as we take a fixed value of the Parefermion's spin the eight vertex model remains integrable. To sum up, the Parafermion weights are on the integrable and critical points of the eight vertex model.

Also it is known that the *Q=0* limit of the Potts model is equivalent to the Abelian Sandpile model (ASM) [29], this means that a Parafermion in the ASM may exists but with $s = 0$, which is a boson rather than fermion. This may explain why the field associated with ASM is a ghost field [30].

5.Acknowledgements

We acknowledge discussions with Y. Ikhlef and M. A. Rajabpour which have been very helpful to us. We are also indebted to Tayeb Jamali who was active in a numerical investigation of these results which has been left for a later publication.